\documentclass[pra,twocolumn,10pt]{revtex4}

\usepackage{amsfonts,amssymb,amsmath,amsthm,xcolor}
\usepackage{bbold,mathtools,braket}
\usepackage{tikz,xcolor,color}	
\usetikzlibrary{calc,backgrounds,fit,decorations.pathreplacing}  
\usepackage{hyperref}

\newcommand{\tracenorm}[1]{{\left  \|  {#1} \right \|_{{\rm tr} }   }}

\DeclareMathOperator{\tr}{Tr}

\def\trhox {\widetilde \rho }
\def\trhoy {\widetilde \rho' }

\def\>{\rangle}

\def\<{\langle}

\begin{document} 

\title{Quantum homomorphic encryption from quantum codes}

\author{Yingkai \surname{Ouyang}}
\affiliation{Singapore University of Technology and Design, 8 Somapah Road, Singapore}
\affiliation{Centre for Quantum Technologies, National University of Singapore, 3 Science Drive 2, Singapore}
\email{yingkai\_ouyang@sutd.edu.sg}

\author{Si-Hui \surname{Tan}}
\affiliation{Singapore University of Technology and Design, 8 Somapah Road, Singapore}
\affiliation{Centre for Quantum Technologies, National University of Singapore, 3 Science Drive 2, Singapore}

\author{Joseph F. \surname{Fitzsimons}  }
\affiliation{Singapore University of Technology and Design, 8 Somapah Road, Singapore}
\affiliation{Centre for Quantum Technologies, National University of Singapore, 3 Science Drive 2, Singapore}

\begin{abstract}
The recent discovery of fully-homomorphic classical encryption schemes has had a dramatic effect on the direction of modern cryptography. Such schemes, however, implicitly rely on the assumptions that solving certain computation problems are intractable. Here we present a quantum encryption scheme which is homomorphic for arbitrary classical and quantum circuits which have at most some constant number of non-Clifford gates. Unlike classical schemes, the security of the scheme we present is information theoretic and hence independent of the computational power of an adversary.
\end{abstract}

\maketitle
 
 \section{Introduction}

Harnessing the power of quantum mechanics to build cryptosystems \cite{BB84,Eke91}
is a key motivation for developing quantum technologies. 
Quantum cryptography often provides information-theoretic security guarantees relying only on the correctness of quantum mechanics, and avoids the need for assumptions about the computational hardness of certain problems as is common place in many classical cryptographic protocols. Such successful quantum approaches to cryptographic problems include secure randomness generation \cite{PAM10,VV12}, coin-flipping \cite{GVW99,ATVY00,Amb01}, secret sharing \cite{HBB99,CGL99,Got00} and bit-commitment \cite{DKSW07,CKe11,kent2011unconditionally,kent2012unconditionally}. One area in particular that has seen significant progress in recent years is the development of quantum cryptographic protocols for delegated computation \cite{dunjko2014composable}, which includes blind quantum computation \cite{BFK09, ABE08, Barz20012012,MFu13, PhysRevLett.111.230501, mantri2013optimal}, and verifiable quantum computation \cite{FK13,RUV13,SFKW13,McK10,HaT15,HaH16}. Homomorphic encryption has been recognised as an important primitive for building secure delegated computation protocols for many decades \cite{RAD78}. It provides a processing functionality for encrypted quantum data which stays secret during the evaluation, and a scheme is said to be \textit{fully-homomorphic} if it allows for arbitrary quantum computation. 
 Despite widespread interest in this problem, it was not until 2009 that the first computationally secure classical scheme for fully homomorphic encryption (FHE) was discovered \cite{Gen09}, with many improvements following rapidly from this initial discovery \cite{DGHV10,GHS12}. Recently this topic also has drawn attention within the quantum information community \cite{Liang13, Liang15, FBS2014, Chi05,PhysRevLett.109.150501, tan2016quantum}.
One might wonder if quantum cryptosystems can offer unconditionally secure homomorphic encryption schemes and whether homomorphic encryption could be extended to allow for evaluation of quantum circuits.

Like their classical counterparts, quantum homomorphic encryption (QHE) schemes comprise of four parts: key generation, encryption, evaluation, and decryption. Unlike blind quantum computation \cite{BFK09}, in which the computation to be performed forms part of the secret, QHE schemes do not have secret circuit evaluations. Rather, they serve to obscure only the information that is contained within the state to be processed using the chosen circuit. The extent to which a scheme is secure depends on its specifics, and in previous work has varied depending on the precise nature of the set of computations which can be performed on the encrypted input. QHE schemes described in Refs.~\cite{PhysRevLett.109.150501, tan2016quantum} offer some information theoretic security, but this is only in the form of a gap between the information accessible with and without the secret key, a notion of security which does not imply the stronger notion of security under composition. These schemes are also limited in the set of operations that can be performed on the encrypted data. 
The scheme in \cite{PhysRevLett.109.150501} only allows computations in the BosonSampling model, while that in \cite{tan2016quantum} is not known to support encoded universal quantum computing. 
Recently Dulek, Schaffner and Speelman \cite{DSS16} 
used the garden-hose model of computation with Broadbent and Jeffery's quantum homomorphic schemes \cite{BJe15} to allow the evaluation of polynomial-depth circuits. 
Several other schemes for computing on encrypted data have previously been introduced which offer universal quantum computation, but require interactions between the client and evaluator \cite{Liang13, Liang15, FBS2014, Chi05}. This requirement for interaction places them outside of the formalism of homomorphic encryption. 

The difficulty in creating a perfectly secure quantum fully-homomorphic encryption (QFHE) scheme persists, and is in line with the no-go results that perfect \cite{YPF14} and approximate \cite{newman2017limitations} information-theoretic security whilst enabling arbitrary processing of encrypted data is impossible, unless the size of the encoding grows exponentially. 
 Nonetheless, given the growing interest in QHE schemes and the multitude of possibilities, Broadbent and Jeffery set out to provide a rigorous framework for defining QHE schemes \cite{BJe15}, basing their security definitions on the requirement for indistinguishability of codewords under chosen plaintext attack. 
Broadbent and Jeffery also require that a quantum {\it fully} homomorphic encryption satisfies two properties: correctness and compactness. Perfect correctness occurs when the evaluated output on the cipherstate after decryption is equivalent to the output of the direct evaluation on the quantum plaintext.
A scheme is compact if the circuit complexity of decryption algorithm does not depend on the computation to be evaluated and scales only polynomially in the size of the plaintext. 
An important implication of the compactness requirement for QHE schemes is that the decryption algorithm of such schemes cannot in any way depend on the evaluated computation. This necessarily implies that a one-time-padding scheme, where random Paulis encrypt the quantum input, does not qualify as a QHE scheme. This is because the decryption algorithm of a one-time-padding scheme is not independent of the evaluated computation.

We present a QHE scheme that supports evaluation of quantum circuits with a constant number of $T$-gates on multiple copies of the input qubits while providing strong information theoretic security guarantees. 
The proposed scheme,
which requires the encoder to be able to produce the multiple copies of the input state,
 builds on constructions taken from quantum codes to provide gates for universal quantum computation. The block of qubits that contains the code is embedded in a larger set of qubits that are initialized in a maximally mixed state. The qubits are then shuffled in a specific but random way to hide the qubits that contain that code.
In our scheme, the evaluation of each $T$-gate succeeds with a probability of half. 
This leads to a trade-off between the size of the encoding and the success probability, since the probability of success can be amplified by encoding several instances of the plaintext in parallel. To achieve a constant success probability, however, the size of this encoding would scale exponentially in the total number of $T$-gates to be performed,
  Hence, in order to maintain compactness, we restrict evaluation to circuits containing at most some constant number of $T$ gates.
The computational model that we consider is non-trivial in the sense that even performing just the Clifford operations on an arbitrary quantum input is known to be hard unless the polynomial hierachy collapses \cite{bravyi2005universal,JozsaNest14,koh2015further,bouland2017quantum}.

Our protocol guarantees that the trace distance between ciphertexts corresponding to arbitrary pairs of quantum inputs is exponentially suppressed in the key size less half the total number of qubits used for the quantum input.
An encryption scheme has entropic security
if an adversary whose min-entropy on the encrypted message is upper
bounded cannot guess any function of the message \cite{Des09,DeD10}.
When the quantum min-entropy of the source in our scheme is sufficiently large,
the trace distance between ciphertexts is exponentially suppressed in only the key size.
Since an exponentially suppressed trace distance implies entropic security \cite{Des09},
our scheme is also secure for high entropy quantum inputs on any number of qubits with a constant key size.

This is a significantly stronger security guarantee than previous homomorphic encryption schemes presented in Ref.~\cite{PhysRevLett.109.150501} and Ref.~\cite{tan2016quantum}. 
Moreover the computation power of our scheme is similar to that of Broadbent and Jeffery's while avoiding reliance on the classical homomorphic encryption scheme. 
This use of classical fully homomorphic encryption is the weakest link in the Broadbent-Jeffery cryptosystem, since it relies on computational assumptions 
\footnote{We note that since the Broadbent-Jeffery cryptosystem utilises the classical FHE to compute sums of hidden subsets, it may be possible to remove the computational assumptions by replacing the FHE scheme with an information theoretically secure scheme which allows evaluation of only linear circuits. However, no analysis of such a modification has yet appeared in the literature.}.
When considering the case of no $T$-gates, it is instructive to compare the decryption complexities of our QHE scheme and the non-QHE quantum one-time padding scheme. In the one-time padding scheme, it is necessary for the decryption routine to take into account a description of the entire circuit which has been performed. Due to the number of Clifford group operations, this implies that the decryption algorithm has complexity at least quadratic in the plaintext size.
In contrast our QHE scheme requires only a linear complexity for decryption, as long as it has at most a constant number of $T$-gates in the evaluated computation.

\begin{figure}
        \includegraphics[width=\columnwidth]{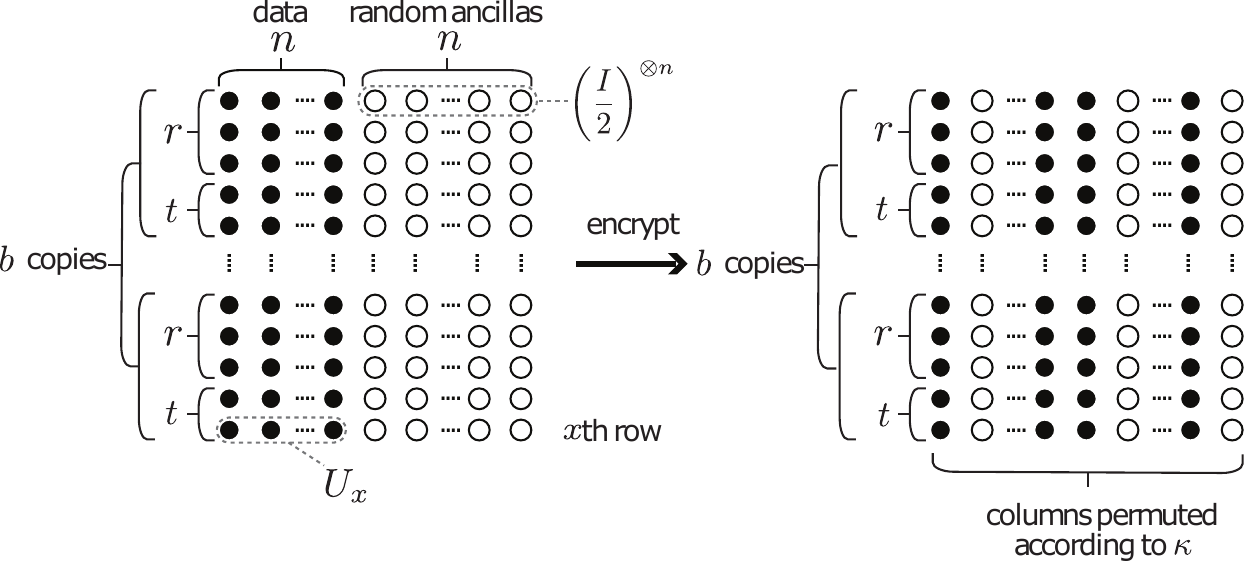}
	\caption{The shaded circles represent data qubits. Within the $x$-th row, the $n$ data qubits are in a code encoded by $U_x$. The unshaded circles are ancilla qubits which are completely mixed. There are $b$ copies of $r$ logical qubits. A random permutation of the columns completes the encryption procedure of our quantum homomorphic encryption scheme.}
	\label{fig:encoding}
\end{figure}

\section{Our QHE scheme}

\begin{figure}
        \includegraphics[width=\columnwidth]{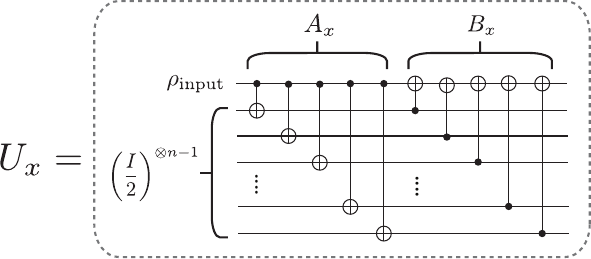}
	\caption{Figure shows the encoding quantum circuit $U_x=B_x A_x$ that is applied on the first $n$ qubits in the $x$-th row. Each line represents one qubit and the gates are applied in the order from left to right.}
	\label{fig:Ux}
\end{figure}

Our QHE scheme takes as its input a $r$-qubit state $\rho_{\rm input}$, and $t$ independent copies of the magic state $|T\> \< T| = \frac{ I }{2}  + \frac{X+Y}{2\sqrt{2}} $, all arranged in a single column (See Figure \ref{fig:encoding}), where $I,X,Y,Z$ are the usual Pauli matrices.
We then introduce $(2n-1)$ more columns of maximally mixed qubits to obtain a grid of qubits with $r+t$ rows and $2n$ columns. 
Here, we require $\frac{n-1}{4}$ to be a nonnegative integer. Of the new columns introduced, $n-1$ of them are incorporated as data qubits while the remaining $n$ columns are used as ancillae in the encryption.
An encoding quantum circuit $U = U_1 \otimes \dots \otimes U_{r+t}$ applies row-wise on the first $n$ columns, where $U_x$ operates on the $x$-th row (see Figure \ref{fig:Ux}). 
We take $A_x$ and $B_x$ to denote the first and last $n-1$ gates in $U_x$ respectively, so that $U_x  =  B_x A_x$.
Applying $U$ spreads the quantum input from just the first column to the first $n$ columns. Since every qubit not residing on the first column is maximally mixed, the encoding circuit on each row encodes the quantum data on the first column into a random quantum code, the resultant quantum information of which resides in a random codespace on the first $n$ columns.
Namely on the $x$-th row, the encoding maps an arbitrary state 
$\rho_{\rm input} = \frac{I+  r_X X + r_Y Y + r_Z Z }{2}$ in the first column and with maximally mixed states on the remaining $n-1$ columns to the mixed state
$U_x \left(
\frac{I+  r_X X + r_Y Y + r_Z Z }{2 } \otimes (( \frac I 2) ^{\otimes n-1}) 
\right) U_x ^\dagger$
which is equivalent to 
$2^{-n} (I ^{\otimes n }+  r_X X ^{\otimes n } + r_Y Y ^{\otimes n }+ r_Z Z ^{\otimes n}) $.
We emphasize at this point that any state in our random codespace is a highly mixed state.
Encryption is then achieved via randomly permuting the $2n$ columns using a secret permutation $\kappa$.
Permuting the columns brings the quantum information to be processed from the first $n$ columns to the columns $k_1, \dots, k_n$, where $1 \le k_1 < \dots < k_n \le 2n$.
For the decryption algorithm, one performs the inverse permutation of the columns $\kappa ^{-1}$, 
followed by the inverse unitary $U ^\dagger$ on the first $n$ columns of the grid.
Finally every qubit in the rows $r+1$ to $r+t$ are measured in the computation basis.
The quantum output of our scheme is then located on the first $r$ rows of the first column of our grid of qubits.

The single qubit logical Clifford operators of each of our random codes on $n$ qubits are transversal gates on those $n$ qubits.
This means that a logical $G$ operator on the $x$-th row is $G ^{\otimes n }$ that operates on the first $n$-columns for every Clifford gate $G$ in the set generated by $\{S,H\}$ where
 $S = |0\>\<0| +  i|1\>\<1|$ and $H = \frac{X+Z}{\sqrt 2}$ is the Hadamard matrix.
To see this, notice that  
$U_x  (Z \otimes I ^{\otimes n-1}) =  
 B_x A_x (Z \otimes I ^{\otimes n-1})= 
 B_x (Z \otimes I ^{\otimes n-1}) A_x= 
Z ^{\otimes n} B_x A_x= 
Z ^{\otimes n} U_x.  $
Hence our encoding circuit $U_x$ maps the physical $Z$ on one qubit to $Z ^{\otimes n}$.
Similarly,
$U_x (X \otimes I ^{\otimes n-1}) =  
 B_x A_x (X \otimes I ^{\otimes n-1})=  
 B_x X ^{\otimes n} A_x =  
X^n \otimes X ^{\otimes n-1}  B_x  A_x .$
Since $n$ is odd, we get 
$U_x (X \otimes I ^{\otimes n-1})  = X ^{\otimes n}	U_x $.
Thus our encoding circuit $U_x$ maps the physical $X$ on one qubit to $X^{\otimes n}$.
Since $Y=iXZ$ and $n-1$ is also divisible by 4, our encoding circuit $U_x$ maps the physical $Y$ on one qubit to $Y^{\otimes n}$.
Now $X ^{\otimes n}$ and $Z ^{\otimes  n}$ anticommute because $n$ is odd, and 
the $Y ^{\otimes n }$ anticommutes with $X ^{\otimes  n}$ and $Z ^{\otimes  n}$.
Upon conjugation by $H ^{\otimes n}$, $X ^{\otimes n}$ becomes $Z ^{\otimes n}$ and vice versa, and $Y ^{\otimes n}$ becomes $-Y ^{\otimes n}$.
Upon conjugation by $S ^{\otimes n}$ gate, $X ^{\otimes n}$ and $Y ^{\otimes n}$ 
become $Y ^{\otimes n}$ and $-X ^{\otimes n}$ respectively. 
Transversality  of the logical CNOT with control and target on distinct rows 
follows immediately from the transversality of the logical $X$ operation. 
Thus the transversal Clifford operations on the $n$ columns containing the encoded quantum data are precisely the logical Clifford operations.

The evaluator operates independently and identically (i.i.d) on not $n$ but $2n$ columns of qubits, $n$ columns of which are the maximally mixed state.
The i.i.d structure of the evaluator's operations allows these operations to commute with any secret permutation of the columns of the qubits on the grid.
In addition, the evaluators' operations necessarily map the $n$ columns of qubits initialized in the maximally mixed state to the maximally mixed state, thereby implementing i.i.d quantum operations on only the columns containing the encoded quantum data.
This allows the evaluator to perform transversal gates on the $n$ columns with the quantum data without knowing where they are located.

The evaluation algorithm takes as input a sequence of unitary operations $(V_1 ,\dots, V_d)$ to be performed securely on $r$ qubits, where each $V_i$ applies either a Clifford gate or a $T$ gate locally on a single qubit, or applies a CNOT locally on a pair of qubits.
The number of $T$-gates to be applied locally amongst the unitary operations $V_1, \dots, V_d$ is at most $t$.
The circuit to be evaluated is $V= V_d \dots V_1$,
where the evaluator applies homomorphisms of the gates $V_1$ to $V_d$ sequentially.

When $V_i$ is a unitary operation that applies a Clifford gate $G$ locally on the $x$-th qubit,
the evaluator can apply the logical $G$-gate on our random code on the $x$-th row without any knowledge of the data columns $k_1, \dots, k_n$.
To do so, the evaluator simply applies the unitary $G ^{\otimes 2n}$ on the $2n$ qubits located on the $x$-th row on each copy. 
Since conjugating a maximally mixed state $\frac{I}{2}$ by any qubit unitary operation yields also a maximally mixed state, 
the net effect is to apply the unitary $G^{\otimes n}$ on the qubits in the encrypted data columns $k_1, \dots, k_n$ on the $x$-th row, which is the logical $G$-gate on the $x$-th row.

When $V_i$ is a unitary operation that applies a CNOT gate with control on the $x$-th qubit and target on the $y$-th qubit, denoted as CNOT$_{x,y}$,
the evaluator can also apply the corresponding logical CNOT gate 
on our random code on the $x$-th and $y$-th row without any knowledge of the data columns $k_1, \dots, k_n$.
To do so, the evaluator simply applies a CNOT with 
control qubit on the $x$-th row and the $j$-th column 
and 
target qubit on the $y$-th row and the $j$-th column 
for every $j = 1,\dots, 2n$. 
As before, the net effect is to apply the unitary CNOT$^{\otimes n}$ on the qubits in the encrypted data columns $k_1, \dots, k_n$ with control qubits on the $x$-th row and target qubits on the $y$-th row, which is the correct logical CNOT-gate, which we denote as $\overline{\rm CNOT}_{x,y}$.
 
\begin{figure}[htb]
  \includegraphics[width=\columnwidth]{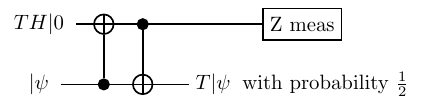} 
    \caption{Gate teleportation of the $T$-gate without correction. \label{fig:gate-teleport}}
    \end{figure}

When $V_i$ is a unitary operation that applies the $k$-th non-Clifford gate $T = |0\>\<0| + e^{i\pi /4}|1\>\<1|$ on the $x$-th qubit,
the evaluator has to perform gate teleportation \cite{GCh99,ZLC00}.
Now consider gate teleportation of a single-qubit gate $T$.
Omitting the correction operation required by gate teleportation allows this procedure to succeed with probability $\frac 1 2$ as depicted in Figure \ref{fig:gate-teleport}.
The principle of deferred measurement \cite{nielsen-chuang} allows deferment of the required measurement until decryption.  
To implement gate teleportation of the logical $T$ operation,
the evaluator applies homomorphisms for CNOT$_{x,r+k}$ 
and CNOT$_{r+k,x}$ sequentially.
Because of the ancilla columns being in the maximally mixed state, 
the unitary $\overline{\rm CNOT}_{x,r+k}$ 
followed by the unitary $\overline{\rm CNOT}_{r+k,x}$ are effectively applied on the data columns $k_1, \dots, k_n$.
For the data qubits encoded on the random codespace, this action implements a logical T gate on the random codespace with probability $\frac{1}{2}$ 
when the outcome of the logical Z-measurement is 0.
This scheme works because by replacing each Pauli in the Pauli decomposition of $|T\>\<T|$ with the corresponding logical Pauli, we obtain precisely the logical $|T\>\<T|$-state.
We emphasize that the outcome of the logical Z-measurement is a flag for the correctness of the implementation of the $T$-gate; if the outcome is 0 the gate is successfully implemented, otherwise the implementation fails.

Our scheme with heralded success satisfies the correctness condition of Broadbent and Jeffery.
Each copy of our scheme yields the correct quantum output with constant probability $2^{-t}$.
Extra copies simply amplify the probability of success.
Thus although each instance of our scheme implements $T$ non-deterministically, it has \textit{heralded} perfect completeness: 
namely, $b = {\lfloor\sqrt{ \frac{\alpha}{2}} + 1\rfloor^2 2^{2t}}$ copies of our scheme yields the correct output in at least one copy with probability at least $1-e^{-\alpha}$, 
and we know which of the $b$ copies yield the correct output.
A large $\alpha$ amplifies the success probability close to unity.

In the three-part algorithm for the decryption, $U^\dag$ requires $2(n-1)b (r+t)$ gates, and unpermuting of the columns requires at most $(2n-1)b(r+t)$ gates (the largest cycle contained in any element of $S_{2n}$ is a $(2n)$-cycle which can be written as a product of $2n-1$ swaps). The remainder of the decryption involves a readout of $Z$ measurements and discarding a subsystem. Since $t$ is constant, $b$ is also constant, and the total number of gates required for decryption scales linearly with $r$ and is independent of the depth of the circuit to be evaluated. 
Hence, our scheme is compact for the family of circuits on $r$ qubits with a constant maximum number of $T$ gates and any number of Clifford gates.  

Randomly permuting the columns of qubits obfuscates the subset of columns where the quantum information resides, thereby encrypting the quantum data.
The maximum trace distance between any two quantum ciphertexts with min-entropy $h$ is 
\begin{align}
\epsilon \le \sqrt{2^{p-h}} \binom{2n}{n}^{-1/2}
, \label{eq:tracedist}
\end{align}  
where $p=b(r+t)$.
In the worst case, where $h=0$, $\epsilon$ is exponentially suppressed in $n$ as long as the key size $n$ grows linearly with the input size $r$. 
However when $t=0,b=1$ and $h =r-x$ for any constant 
$x$, the key size $n$ can be independent of the input size $r$ while having $\epsilon$ being exponentially suppressed in $n$. 
In both cases, any two quantum ciphertexts are essentially the maximally mixed state and hence indistinguishable in our scheme.

To obtain Eq.~(\ref{eq:tracedist}), we first obtain a Pauli decomposition of any arbitrary state that the evaluator receives.
Let the density matrices $\rho$ and $\rho'$ on $b$ copies of $2n(r+t)$ qubits be any two arbitrary inputs to the scheme before encoding and encryption.
Let $\widetilde \rho$ and $\widetilde \rho'$ be the corresponding states after encoding and encryption. 
Then $\epsilon = \frac 1 2 \max_{\rho, \rho'} \|\widetilde \rho - \widetilde \rho' \|_{\rm tr}$.
In this maximization, 
only the $p = b(r+t)$ qubits in the first column are arbitrary, and the remaining columns are in the maximally mixed state.
Note that $\tracenorm{ \widetilde \rho - \widetilde \rho' }= \tr ( M ( \trhox  -\trhoy ) )$ 
for some optimal Hermitian $M$ diagonal in the same basis as 
$ \widetilde \rho - \widetilde \rho' $, with 
eigenvalues equal to +1 or -1.
More precisely, if $\trhox - \trhoy$ has the spectral decomposition\
$\sum_{i } \lambda_i |i\>\<i|$, then
$M = \sum_{i} {\rm sign}(\lambda_i) |i\>\<i|$, where ${\rm sign}(\lambda_i) = 1$ if $\lambda_i \ge 0$ and ${\rm sign}(\lambda_i) = -1$ otherwise.
Now define $\sigma_0 = I, \sigma_1 = X, \sigma_2 = Y, $ and $\sigma_3 =Z$.
Let $\mathcal M_{p,2n}(\mathbb Z_4)$ denote the set of all matrices with $p$ rows and $2n$ columns and entries from $\{0,1,2,3\}$.
Given any matrix $A \in \mathcal M_{p,2n}(\mathbb Z_4)$,
let $a_{x,y}$ denote its component in the $x$-th row and the $y$-th column.
Define the unitary matrix $\sigma_A$ to be one that applies $\sigma_{a_{x,y}}$ on the $x$-th row and $y$-th column of our grid of qubits for every $x= 1,\dots, p$ and $y = 1,\dots , 2n$.
Define the set of all column permutations of $\sigma_A$ as $S_A$,
and the corresponding symmetric sum of $\sigma_A$ as 
$\widetilde \sigma_{A}  = \sum_{\tau \in 	 S_A} \tau.$
Let $\Omega$ denote the set of non-zero column vectors of length $p$ with entries from $\{0,1,2,3\}$.
For all ${\bf v} \in \Omega$, let $\varphi({\bf v})$ denote a matrix with $p$ rows and $2n$ columns such that its first $n$ columns are identical to ${\bf v}$ and the last $n$ columns have all entries equal to zero.
Notice that for distinct ${\bf v},{\bf v}' \in \Omega$, $\widetilde \sigma_{\varphi({\bf v})}$ and $\widetilde \sigma_{\varphi({\bf v}')}$ are also distinct.
Let $\mathcal S$ denote some minimal subset of $\mathcal M_{p,2n}(\mathbb Z_4)$ such that 
$\{ \widetilde \sigma_A :A \in \mathcal S \} = \{ \widetilde \sigma_A :A \in \mathcal M_{p,2n}(\mathbb Z_4) \}$.
Now we can always have $\varphi({\bf v}) \in \mathcal S$ for every ${\bf v} \in \Omega$.
Let $\widetilde M = \frac{1 }{ (2n)! } \sum_{  \pi }	  \pi   M  \pi ^\dagger$, where $\pi$ is any column permutation.
Then we can write $\widetilde M = \sum_{ A \in \mathcal S } a_A \widetilde \sigma_A$, for appropriate real constants $a_A$.

Linearity and the cyclic property of the trace give 
$\tr ( M ( \widetilde \rho  -\widetilde\rho' ) )  
=
\tr ( \widetilde M (   \rho  - \rho' ) )  $.  
Using the decomposition
$
\rho  - \rho' = 
\sum_{ {\bf v} \in \Omega } 
	\frac{ r_{\bf v}  -  r'_{\bf v} }  { 2^{2n p } } \sigma_{\varphi({\bf v})}
	$
for appropriate real constants $r_{\bf v}$ and $r'_{\bf v}$,
the decomposition of $\widetilde M$, the linearity of trace, and the triangle inequality, we get
\begin{align}
\tracenorm{ \widetilde \rho - \widetilde \rho' }
&\le 
\sum_{ 	{\bf v} \in \Omega } \sum_{	B \in \mathcal S}
\left| \tr
a_{B} \widetilde \sigma_{B} \frac{(r_{\bf v} - r'_{\bf v})}{ 2^{2n p }}  \sigma_{\varphi({\bf v})}  \right|. 
\end{align}
Orthogonality of the Pauli operators under the Hilbert-Schmidt inner product gives
\begin{align}
\tracenorm{ \widetilde \rho - \widetilde \rho' }
& \le 
\sum_{ 	{\bf v} \in \Omega} 
\left| \tr
a_{\varphi({\bf v})}  (r_{{\bf v}} - r'_{ {\bf v}})     \right|. 
\label{eq:rho-diff}
\end{align}
The Cauchy-Schwarz inequality implies that 
$
\|\widetilde{\rho}-\widetilde{\rho}'\|_{\rm tr}
\le
\sqrt{ \sum_{{\bf v \in \Omega}} a_{\varphi({\bf v})}^2 }
\sqrt{ \sum_{{\bf v \in \Omega}}(r_{\bf v}-r'_{\bf v})^2 }$.
Since in Loewner order $\widetilde M^2 \leq I$, and hence $\tr (\widetilde M^2) \le \tr(I)$, we have $\sum_{{\bf v} \in \Omega} a_{\varphi({\bf v})}^2  \binom{2n}{n} \leq 1$.
Next we show that $ \sum_{{\bf v \in \Omega}}(r_{\bf v}-r'_{\bf v})^2  \le 2^{p-h+2}$ if the $p$-qubit inputs to the first column of our scheme has a quantum min-entropy \cite{KRS2009} of $h$.
Let $\tau$ and $\tau'$ be unencrypted $p$-qubit states, with $\rho = U (\tau \otimes (I/2)^{\otimes (2n-1)p}) U ^\dagger$
and $\rho' = U (\tau' \otimes (I/2)^{\otimes (2n-1)p}) U ^\dagger$.
The Pauli decompositions 
$\tau = 2^{-p}\sum_{{\bf v} \in \Omega} r_{\bf v} \sigma_{\bf v}$ 
and 
$\tau' = 2^{-p}\sum_{{\bf v} \in \Omega} r_{\bf v} \sigma_{\bf v}$
imply that 
 $\tr((\tau-\tau')^2)  = \sum_{{\bf v}\in \Omega} (r_{\bf v} - r'_{\bf v})^2 2^{-p}.$
Given the min-entropy of $\tau$ and $\tau'$, 
their maximum eigenvalue is $2^{-h}$.
Hence $\tr((\tau-\tau')^2) \le 2^{-h+2}$.
 Then $\sum_{{\bf v}\in \Omega} (r_{\bf v} - r'_{\bf v})^2 \le 2^{p-h+2}$ and 
 Eq.~(\ref{eq:tracedist}) can thereby be obtained.


\section{Discussions}

In summary, our QHE scheme encodes the quantum input using random codes, encrypts and decrypts via a secret permutation, and allows the evaluator to compute a constant number of non-Clifford ($T$) gates on the encrypted data. Since the encrypted quantum ciphertexts are almost indistinguishable, the evaluator is essentially oblivious to the quantum input, which gives our scheme its information-theoretic security. 
Moreover, our scheme trivially allows homomorphisms  
of arbitrary reversible linear boolean circuits using the homomorphisms of CNOT and $X$ gates.
Our scheme may also offer robustness to noise when the encryptor holds purifications to the maximally mixed states used in the random encodings and performs a recovery operation dependent on the classical measurement outcomes on her ancillary registers, and we leave this for future study.
We also like to point out that it is sometimes preferable to use a quantum one-time padding scheme as opposed to our scheme for delegated computation, for example when no $T$ gates need to be performed and when all the input states are stabilized by Clifford gates.
 
 \section{Acknowledgements}

The authors thank Anne Broadbent and Stacey Jeffery for useful correspondence about quantum homomorphic encryption. JFF also thanks Renato Renner for useful discussions relating to security definitions. This material is based on research supported in
part by the Singapore National Research Foundation under NRF Award No. NRF-NRFF2013-01.
JFF and ST acknowledges support from the Air Force Office of Scientific Research under AOARD grant FA2386-15-1-4082.

\bibliography{../../../mybib}{}

\begin{thebibliography}{10}

\bibitem{BB84}
C.~H. Bennett and G.~Brassard, ``{Quantum cryptography: Public key distribution
  and coin tossing},'' in {\em Proceedings of IEEE International Conference on
  Computers, Systems and Signal Processing}, vol.~175, New York, 1984.

\bibitem{Eke91}
A.~K. Ekert, ``Quantum cryptography based on {B}ell's theorem,'' {\em Phys.
  Rev. Lett.}, vol.~67, pp.~661--663, Aug 1991.

\bibitem{PAM10}
S.~Pironio, A.~Ac\'in, S.~Massar, A.~B. de~la Giroday, D.~N. Matsukevich,
  P.~Maunz, S.~Olmschenk, D.~Hayes, L.~Luo, T.~A. Manning, and C.~Monroe,
  ``Random numbers certified by {B}ell's theorem,'' {\em Nature}, vol.~464,
  no.~7291, pp.~1022--1024, 2010.

\bibitem{VV12}
U.~Vazirani and T.~Vidick, ``Certifiable quantum dice: Or, true random number
  generation secure against quantum adversaries,'' in {\em Proceedings of the
  Forty-fourth Annual ACM Symposium on Theory of Computing}, STOC '12, (New
  York, NY, USA), pp.~61--76, ACM, 2012.

\bibitem{GVW99}
L.~Goldenberg, L.~Vaidman, and S.~Wiesner, ``Quantum gambling,'' {\em Phys.
  Rev. Lett.}, vol.~82, pp.~3356--3359, Apr 1999.

\bibitem{ATVY00}
D.~Aharonov, A.~Ta-Shma, U.~V. Vazirani, and A.~C. Yao, ``Quantum bit escrow,''
  in {\em Proceedings of the Thirty-second Annual ACM Symposium on Theory of
  Computing}, STOC, (New York, NY, USA), pp.~705--714, ACM, 2000.

\bibitem{Amb01}
A.~Ambainis, ``A new protocol and lower bounds for quantum coin flipping,'' in
  {\em Proceedings of the Thirty-third Annual {ACM} Symposium on Theory of
  Computing}, STOC '01, (New York, NY, USA), pp.~134--142, ACM, 2001.

\bibitem{HBB99}
M.~Hillery, V.~Bu\ifmmode~\check{z}\else \v{z}\fi{}ek, and A.~Berthiaume,
  ``Quantum secret sharing,'' {\em Phys. Rev. A}, vol.~59, pp.~1829--1834, Mar
  1999.

\bibitem{CGL99}
R.~Cleve, D.~Gottesman, and H.-K. Lo, ``How to share a quantum secret,'' {\em
  Phys. Rev. Lett.}, vol.~83, pp.~648--651, Jul 1999.

\bibitem{Got00}
D.~Gottesman, ``Theory of quantum secret sharing,'' {\em Phys. Rev. A},
  vol.~61, p.~042311, Mar 2000.

\bibitem{DKSW07}
G.~M. D'Ariano, D.~Kretschmann, D.~Schlingemann, and R.~F. Werner,
  ``Reexamination of quantum bit commitment: The possible and the impossible,''
  {\em Phys. Rev. A}, vol.~76, p.~032328, Sep 2007.

\bibitem{CKe11}
A.~Chailloux and I.~Kerenidis, ``Optimal bounds for quantum bit commitment,''
  in {\em Foundations of Computer Science (FOCS), 2011 IEEE 52nd Annual
  Symposium on}, pp.~354--362, Oct 2011.

\bibitem{kent2011unconditionally}
A.~Kent, ``Unconditionally secure bit commitment with flying qudits,'' {\em New
  Journal of Physics}, vol.~13, no.~11, p.~113015, 2011.

\bibitem{kent2012unconditionally}
A.~Kent, ``Unconditionally secure bit commitment by transmitting measurement
  outcomes,'' {\em Physical review letters}, vol.~109, no.~13, p.~130501, 2012.

\bibitem{dunjko2014composable}
V.~Dunjko, J.~F. Fitzsimons, C.~Portmann, and R.~Renner, ``Composable security
  of delegated quantum computation,'' in {\em Advances in Cryptology--ASIACRYPT
  2014}, pp.~406--425, Springer, 2014.

\bibitem{BFK09}
A.~Broadbent, J.~Fitzsimons, and E.~Kashefi, ``Universal blind quantum
  computation,'' in {\em Foundations of Computer Science, 2009. FOCS '09. 50th
  Annual IEEE Symposium on}, pp.~517--526, Oct 2009.

\bibitem{ABE08}
D.~Aharonov, M.~Ben-or, and E.~Eban, ``Interactive proofs for quantum
  computations,'' {\em arXiv:0810.5375}, 2008.

\bibitem{Barz20012012}
S.~Barz, E.~Kashefi, A.~Broadbent, J.~F. Fitzsimons, A.~Zeilinger, and
  P.~Walther, ``Demonstration of blind quantum computing,'' {\em Science},
  vol.~335, no.~6066, pp.~303--308, 2012.

\bibitem{MFu13}
T.~Morimae and K.~Fujii, ``Blind quantum computation protocol in which {A}lice
  only makes measurements,'' {\em Phys. Rev. A}, vol.~87, p.~050301, May 2013.

\bibitem{PhysRevLett.111.230501}
V.~Giovannetti, L.~Maccone, T.~Morimae, and T.~G. Rudolph, ``Efficient
  universal blind quantum computation,'' {\em Phys. Rev. Lett.}, vol.~111,
  p.~230501, Dec 2013.

\bibitem{mantri2013optimal}
A.~Mantri, C.~A. P{\'e}rez-{D}elgado, and J.~F. Fitzsimons, ``Optimal blind
  quantum computation,'' {\em Phys. Rev. Lett.}, vol.~111, no.~23, p.~230502,
  2013.

\bibitem{FK13}
J.~F. Fitzsimons and E.~Kashefi, ``Unconditionally verifiable blind
  computation,'' {\em arXiv prnote arXiv:1203.5217}, 2013.

\bibitem{RUV13}
B.~W. Reichardt, F.~Unger, and U.~Vazirani, ``Classical command of quantum
  systems,'' {\em Nature}, vol.~496, pp.~456--460, 04 2013.

\bibitem{SFKW13}
S.~Barz, J.~F. Fitzsimons, E.~Kashefi, and P.~Walther, ``Experimental
  verification of quantum computation,'' {\em Nat. Phys.}, vol.~9,
  pp.~727---731, 11 2013.

\bibitem{McK10}
M.~McKague, ``Self-testing graph states,'' in {\em Theory of Quantum
  Computation, Communication, and Cryptography} (D.~Bacon, M.~Martin-Delgado,
  and M.~Roetteler, eds.), vol.~6745 of {\em Lecture Notes in Computer
  Science}, pp.~104--120, Springer Berlin Heidelberg, 2014.

\bibitem{HaT15}
M.~Hayashi and T.~Morimae, ``Verifiable measurement-only blind quantum
  computing with stabilizer testing,'' {\em Phys. Rev. Lett.}, vol.~115,
  p.~220502, Nov 2015.

\bibitem{HaH16}
M.~Hayashi and M.~Hajdusek, ``Self-guaranteed measurement-based quantum
  computation,'' {\em arXiv preprint arXiv:1603.02195}, 2016.

\bibitem{RAD78}
R.~L. Rivest, L.~Adleman, and M.~L. Dertouzos, ``On data banks and privacy
  homomorphisms,'' {\em Foundations of secure computation}, vol.~4, no.~11,
  pp.~169--180, 1978.

\bibitem{Gen09}
C.~Gentry, ``Fully homomorphic encryption using ideal lattices,'' in {\em
  Proceedings of the Forty-first Annual ACM Symposium on Theory of Computing},
  STOC '09, (New York, NY, USA), pp.~169--178, ACM, 2009.

\bibitem{DGHV10}
M.~Van~Dijk, C.~Gentry, S.~Halevi, and V.~Vaikuntanathan, ``Fully homomorphic
  encryption over the integers,'' in {\em Advances in cryptology--EUROCRYPT
  2010}, pp.~24--43, Springer, 2010.

\bibitem{GHS12}
C.~Gentry, S.~Halevi, and N.~Smart, ``Fully homomorphic encryption with polylog
  overhead,'' in {\em Advances in Cryptology – EUROCRYPT 2012}
  (D.~Pointcheval and T.~Johansson, eds.), vol.~7237 of {\em Lecture Notes in
  Computer Science}, pp.~465--482, Springer Berlin Heidelberg, 2012.

\bibitem{Liang13}
M.~Liang, ``Symmetric quantum fully homomorphic encryption with perfect
  security,'' {\em Quantum Information Processing}, vol.~12, no.~12,
  pp.~3675--3687, 2013.

\bibitem{Liang15}
M.~Liang, ``Quantum fully homomorphic encryption scheme based on universal
  quantum circuit,'' {\em Quantum Information Processing}, pp.~1--11, 2015.

\bibitem{FBS2014}
K.~A.~G. Fisher, A.~Broadbent, L.~K. Shalm, Z.~Yan, J.~Lavoie, R.~Prevedel,
  T.~Jennewein, and K.~J. Resch, ``Quantum computing on encrypted data,'' {\em
  Nat. Commun.}, vol.~5, 01 2014.

\bibitem{Chi05}
A.~M. Childs, ``Secure assisted quantum computation,'' {\em Quantum Info.
  Comput.}, vol.~5, pp.~456--466, Sept. 2005.

\bibitem{PhysRevLett.109.150501}
P.~P. Rohde, J.~F. Fitzsimons, and A.~Gilchrist, ``Quantum walks with encrypted
  data,'' {\em Phys. Rev. Lett.}, vol.~109, p.~150501, Oct 2012.

\bibitem{tan2016quantum}
S.-H. Tan, J.~A. Kettlewell, Y.~Ouyang, L.~Chen, and J.~F. Fitzsimons, ``A
  quantum approach to homomorphic encryption,'' {\em Scientific Reports},
  vol.~6, p.~33467, 2016.

\bibitem{DSS16}
Y.~Dulek, C.~Schaffner, and F.~Speelman, ``Quantum homomorphic encryption for
  polynomial-sized circuits,'' pp.~3--32, 2016.

\bibitem{BJe15}
A.~Broadbent and S.~Jeffery, ``Quantum homomorphic encryption for circuits of
  low t-gate complexity,'' in {\em Annual Cryptology Conference}, pp.~609--629,
  Springer, 2015.

\bibitem{YPF14}
L.~Yu, C.~A. {P{{\'e}}rez-{D}elgado}, and J.~F. Fitzsimons, ``Limitations on
  information-theoretically-secure quantum homomorphic encryption,'' {\em Phys.
  Rev. A}, vol.~90, p.~050303, Nov 2014.

\bibitem{newman2017limitations}
M.~Newman and Y.~Shi, ``Limitations on transversal computation through quantum
  homomorphic encryption,'' {\em arXiv preprint arXiv:1704.07798}, 2017.

\bibitem{bravyi2005universal}
S.~Bravyi and A.~Kitaev, ``Universal quantum computation with ideal clifford
  gates and noisy ancillas,'' {\em Physical Review A}, vol.~71, no.~2,
  p.~022316, 2005.

\bibitem{JozsaNest14}
R.~JOZSA and M.~V.~D. NEST, ``Classical simulation complexity of extended
  clifford circuits,'' {\em Quantum Information and Computation}, vol.~14,
  pp.~0633--0648.

\bibitem{koh2015further}
D.~E. Koh, ``Further extensions of clifford circuits and their classical
  simulation complexities,'' {\em Quantum Information and Computation},
  vol.~17, pp.~0262--0282.

\bibitem{bouland2017quantum}
A.~Bouland, J.~F. Fitzsimons, and D.~E. Koh, ``Complexity classification of
  conjugated clifford circuits,'' {\em arXiv preprint arXiv:1709.01805}, 2017.

\bibitem{Des09}
S.~Desrosiers, ``Entropic security in quantum cryptography,'' {\em Quantum
  Information Processing}, vol.~8, no.~4, pp.~331--345, 2009.

\bibitem{DeD10}
S.~Desrosiers and F.~Dupuis, ``Quantum entropic security and approximate
  quantum encryption,'' {\em Information Theory, IEEE Transactions on},
  vol.~56, pp.~3455--3464, July 2010.

\bibitem{GCh99}
D.~Gottesman and I.~L. Chuang, ``Demonstrating the viability of universal
  quantum computation using teleportation and single-qubit operations,'' {\em
  Nature}, vol.~402, pp.~390--393, 1999.

\bibitem{ZLC00}
X.~Zhou, D.~W. Leung, and I.~L. Chuang, ``Methodology for quantum logic gate
  construction,'' {\em Phys. Rev. A}, vol.~62, p.~052316, Oct 2000.

\bibitem{nielsen-chuang}
M.~A. Nielsen and I.~L. Chuang, {\em {Quantum Computation and Quantum
  Information}}.
\newblock Cambridge University Press, second~ed., 2000.

\bibitem{KRS2009}
R.~Konig, R.~Renner, and C.~Schaffner, ``The operational meaning of min- and
  max-entropy,'' {\em IEEE Transactions on Information Theory}, vol.~55,
  pp.~4337--4347, Sept 2009.

\end{thebibliography}

\bibliographystyle{ieeetr}

\appendix
\section{Security proof}
Here, we provide a detailed proof of Eq.~(\ref{eq:tracedist}). We begin by reviewing some terminology.

In our scheme, qubits are arranged on a grid with $p$ rows and $2n$ columns.
In the encryption procedure, the columns of $2n$ qubits are randomly permuted. 
Hence we consider $S_{2n}$, a symmetric group of order $2n$, and its representation $\nu_{p,2n}$.
For every permutation $\pi\in S_{2n}$, and every $A = \sum_{x=1}^p \sum_{y=1}^{2n} a_{x,y} |x\>\<y|\in \mathcal{M}_{p,2n}$, 
we let $\nu_{p,q} : S_{2n} \to \mathcal M(\mathbb C^{2np})$ be a representation of $S_{2n}$ such that 
for every matrix representation $P_{\pi} = \nu_{p,q}(\pi)$ of $\pi \in S_{2n}$, we have 
\begin{align}
P_\pi \sigma_{A} P_\pi^\dag=\bigotimes_{y=1}^{2n}\left(\bigotimes_{y=1}^p\sigma_{a_{x,\pi(y)}}\right ).
\end{align}
The matrices $P_\pi$ are the permutation operations that permute the columns in our scheme.
With these permutation operations, we can define the set of Paulis generated from all possible column permutations of a particular Pauli $\sigma_A$, given by 
\begin{align}
S_{A}=\left\{P_{\pi} \sigma_{A}P_{\pi}^\dag:    \pi \in S_{2n}\right \} .
\end{align}
The symmetrized Pauli associated with the Pauli $\sigma_A$ is the sum of all the terms in $S_A$ given explicitly by  $\widetilde{\sigma}_{A}=\sum_{\tau\in S_A} \tau.$

Eq.~(1) provides an upper bound on the trace norm of the difference between two encrypted inputs to our scheme, given by $\widetilde \rho$ and $\widetilde {\rho}'$ respectively.
Here $\widetilde \rho$ and $\widetilde {\rho}'$ are uniform mixtures of all column permutations of the unencrypted input $\rho$ and $\rho'$ respectively, where
\begin{align}
\widetilde \rho = \frac{1}{(2n)!} \sum_{\substack{ \pi \in S_{2n}}} P_{\pi} \rho P_{\pi} ^\dagger, 
\quad
\widetilde \rho' = \frac{1}{(2n)!} \sum_{\substack{  \pi \in S_{2n}}} P_{\pi}  \rho' P_{\pi} ^\dagger.
\end{align}
The matrix $\widetilde{\rho}-\widetilde{\rho}'$ admits the spectral decomposition 
\begin{align}
\widetilde{\rho}-\widetilde{\rho}'=\sum_i \lambda_i\ket{\psi_i}\bra{\psi_i} \ ,
\end{align}
where $\{|\psi_i\>\}$ is an eigenbasis of $\widetilde{\rho}-\widetilde{\rho}'$. 
Now let $M=\sum_i {\rm sgn}(\lambda_i)\ket{\psi_i}\bra{\psi_i}$,
where ${\rm sgn}(x) = 1$ if $x\ge 0$ and  ${\rm sgn}(x) = -1$ if $x < 0$.
From the definition of the trace norm, we have
$\|\widetilde{\rho}-\widetilde{\rho}'\|_{\rm tr}= 
\tr\left(M(\widetilde{\rho}-\widetilde{\rho}')\right)$ because 
\begin{align}
\|\widetilde{\rho}-\widetilde{\rho}'\|_{\rm tr}=& \tr\left|\widetilde{\rho}-\widetilde{\rho}'\right |\notag\\
&=\tr\left(\sqrt{\left(\widetilde{\rho}-\widetilde{\rho}'\right)^2}\right )\notag\\
&=\sum_i|\lambda_i|\notag\\
&=\sum_i {\rm sgn}(\lambda_i) \lambda_i\notag\\
&=\tr\left(M(\widetilde{\rho}-\widetilde{\rho}')\right).
\end{align}
The trace norm is non-negative, and hence equal to its absolute value.
Thus,
\begin{align}
\|\widetilde{\rho}-\widetilde{\rho}'\|_{\rm tr}=&\left|\tr\left(M(\widetilde{\rho}-\widetilde{\rho}')\right )\right |,
\end{align}
and using the cyclic property of the trace, we get
\begin{align}
\|\widetilde{\rho}-\widetilde{\rho}'\|_{\rm tr}=&\left|\tr\left(\widetilde{M}(\rho-\rho')\right )\right |, \label{eq:001}
\end{align}
where
\begin{align}
\widetilde M = \frac{1}{(2n)!} \sum_{\substack{  \pi \in S_{2n}}} P_{\pi} M P_{\pi}^\dagger.
\end{align}•
The decomposition of $\widetilde{M}$ into the symmetrized Paulis and the decomposition of the traceless quantity $\rho-\rho'$ into the usual Paulis can be subsituted into Eq.~(\ref{eq:001}) to yield
\begin{align}
\|\widetilde{\rho}-\widetilde{\rho}'\|_{\rm tr}=&\left|\tr\left(\sum_{A\in S}a_A \widetilde{\sigma}_A\sum_{{\bf v}\in \Omega}\frac{r_{\bf v}-r'_{\bf v}}{2^{2np}}\sigma_{\varphi({\bf v})}\right )\right| \label{eq:002}.
\end{align}
Recall that $\Omega$ is the set of all non-zero column vectors of length $p$ with components from the set $\{0,1,2,3\}$, and for every ${\bf v} \in \Omega$, $\varphi({\bf v})$ is a matrix with $2n$ columns where the first $n$ columns are identical to {\bf v} and the remaining $n$ columns are zero vectors.
Using the orthogonality of the Paulis on Eq.~(\ref{eq:002}) yields 
\begin{align}
\|\widetilde{\rho}-\widetilde{\rho}'\|_{\rm tr}
=&
\left|\tr\left( 
\sum_{{\bf v}\in \Omega}a_{\varphi({\bf v})} \frac{r_{\bf v}-r'_{\bf v}}{2^{2np}}
\sigma_{\varphi({\bf v})}^2 
\right )\right|  \notag\\
=&
\left| 
\sum_{{\bf v}\in \Omega}a_{\varphi({\bf v})}(r_{\bf v}-r'_{\bf v}) 
\right| .
\end{align}
Applying the Cauchy-Schwarz inequality on the above yields
\begin{align}
\|\widetilde{\rho}-\widetilde{\rho}'\|_{\rm tr}
\le
\sqrt{ \sum_{{\bf v \in \Omega}} a_{\varphi({\bf v})}^2 }
\sqrt{ \sum_{{\bf v \in \Omega}}(r_{\bf v}-r'_{\bf v})^2 }.
\end{align}
Now define the input states on only the first column of qubits to be
\begin{align}
\tau = \frac{ I ^{\otimes p} + \sum_{{\bf v} \in \Omega} r_{\bf v} \sigma_{\bf v}} {2^p}
, \quad
\tau'  = \frac{ I ^{\otimes p} + \sum_{{\bf v} \in \Omega} r_{\bf v}' \sigma_{\bf v}} {2^p}.
\end{align}
The maximum eigenvalue of each of these states is $2^{-h}$, where $h$ is their min-entropies.
We can use these states to obtain an upper bound on 
$ \sum_{{\bf v \in \Omega}}(r_{\bf v}-r'_{\bf v})^2 $.
Note that
\begin{align}
\tr( (\tau - \tau')^2)
 \le 
\| \tau - \tau' \|_{\rm tr} \| \tau - \tau' \|_\infty,
\end{align}
where $ \| \tau - \tau' \|_\infty$ denotes the $\infty$-norm on the eigenvalues of $ \tau - \tau'$.
Since $\| \tau - \tau' \|_{\rm tr} \le \|\tau\|_{\rm tr} + \| \tau' \|_{\rm tr} \le 1+1 =2 $ and 
$\| \tau - \tau' \|_\infty \le \| \tau \|_{\infty} + \| \tau' \|_\infty = 2^{-h} + 2^{-h} = 2^{-h+1}$, 
we get
\begin{align}
\tr( (\tau - \tau')^2) \le 2^{-h+2}.
\end{align}
Note also that by the orthogonality of the Pauli operators,
\begin{align}
\tr( (\tau - \tau')^2)
&= 
\tr 
\left(
\left(
\sum_{ {\bf v} \in \Omega}
(r_{\bf v} - r_{\bf v}') 2^{-p} \sigma_{\bf v} 
\right) \right.\notag\\
&\left.
\quad\times
\left(
\sum_{ {\bf w} \in \Omega}
(r_{\bf w} - r_{\bf w}')  2^{-p} \sigma_{\bf w} 
\right)
\right)
\notag\\
&=
\tr 
\left(
\sum_{ {\bf v} \in \Omega}
(r_{\bf v} - r_{\bf v}')^2 2^{-2p} \sigma_{\bf v} ^2
\right) 
\notag\\
&= 
\sum_{ {\bf v} \in \Omega}
(r_{\bf v} - r_{\bf v}')^2 2^{-p} .
\end{align}
Hence 
\begin{align}
\sqrt{ \sum_{{\bf v \in \Omega}}(r_{\bf v}-r'_{\bf v})^2 }
\le  2 \sqrt{ 2^{p-h} }.
\end{align}•

To obtain an upper bound for 
$\sqrt{\sum_{{\bf v} \in \Omega} a_{\varphi({\bf v})}^2 }$, 
we obtain upper and lower bounds on $\tr\left(\widetilde{M}^2\right)$.
Now we obtain an upper bound for $\tr\left(\widetilde{M}^2\right).$
By H\"older's inequality, 
$\tr\left(\widetilde{M}^2\right) \le
\| \widetilde M \|_{\rm tr} \| \widetilde M \|_\infty $.
Convexity of the norms then implies that 
\begin{align}
\tr\left(\widetilde{M}^2\right) 
&\le
\|   M \|_{\rm tr} \| M \|_\infty = 2^{2np}  .\label{eq:4}
\end{align}  
The lower bound on $\tr\left(\widetilde{M}^2\right)$ requires us to expand $\widetilde{M}$ in terms of the symmetrized Paulis. Then
\begin{align}
\tr\left(\widetilde{M}^2\right )
= 
\tr\left(\sum_{A,A'\in S}a_A a_{A'}\widetilde{\sigma}_A \widetilde{\sigma}_{A'}\right ) .
\end{align}
By the orthogonality of the symmetrized Paulis and linearity of the trace, we get
\begin{align}
\tr\left(\widetilde{M}^2\right )
=
&\tr\left(\sum_{A\in S} a_A^2(\widetilde{\sigma}_A) ^2 \right )   \notag\\
=&
\sum_{A\in S} a_A^2  \tr\left((\widetilde{\sigma}_A) ^2 \right )  \notag\\
\ge&
\sum_{{\bf v} \in \Omega} a_{\varphi({\bf v})}^2  \tr\left((\widetilde{\sigma}_{\varphi({\bf v})}) ^2 \right ) .
\end{align} 
Now $\widetilde{\sigma}_{\varphi({\bf v})}$ is the sum of $\binom {2n}n $ Pauli operators, 
because there are $\binom {2n}n$ ways to permute the $2n$ columns of a matrix with $n$ identical columns and $n$ columns of zeros. 
Thus $\tr\left(\widetilde{\sigma}_{\varphi({\bf v})}^2\right )  = \binom{2n}{n} 2^{np}$,
and 
\begin{align}
\tr\left(\widetilde{M}^2\right )
\ge & 
\sum_{{\bf v} \in \Omega} a_{\varphi({\bf v})}^2 
\binom {2n}{n} 2^{2np}. \label{eq:5}
\end{align}
Eqs.~(\ref{eq:4}) and (\ref{eq:5}) together imply that  
\begin{align}
\sqrt{ \sum_{{\bf v} \in \Omega} a_{\varphi({\bf v})}^2} \le \binom {2n}{n}^{-1/2} \ .\label{eq:7}
\end{align}
Hence 
\begin{align}
\| \widetilde \rho - \widetilde \rho' \|_{\rm tr}
\le 
  2 \sqrt{ 2^{p-h} } \binom {2n}{n}^{-1/2} .
\end{align}  
The trace distance between two states is half of the trace norm of the difference between the two states, and hence $\epsilon \le \sqrt{ 2^{p-h} } \binom {2n}{n}^{-1/2} .$

\end{document}